\documentstyle[11pt,aaspp4]{article}
\begin{document}

\def\solar{\ifmmode_{\mof d\odot}\;\else$_{\mathord\odot}\;$\fi}

\noindent
{\it Conference Highlights}

\begin{center}

\title{\large \bf The Outer Edges of Dwarf Irregular Galaxies: 
Stars and Gas\footnote{Workshop held in Flagstaff, AZ, USA
on October 10--11, 2002. Proceedings are available on-line
at http://www.lowell.edu/Workshops/Lowell02/ .}}

\end{center}

\medskip

\section{Introduction}

The outer edges of dwarf irregular galaxies have been an area of 
interest for about 40
years. Its origin can be traced back at least to Sandage's (1962) 
discussion of the outer
parts of IC 1613, which, on Baade's deep red photographic plates, 
showed an envelope of
faint red stars that extended outwards significantly beyond the 
main parts of the galaxy.
Scattered reports of the properties of other dwarf galaxies followed 
at a low rate until the
recent work of Hunter and her collaborators highlighted the importance 
of the edges of
irregular galaxies, both because of their importance in illuminating 
the evolution of the
galaxies and because of the sensitivity of their properties to 
the nature of the intergalactic
medium. Recent research in this area has flourished, as the Workshop clearly
demonstrated.

\section{The Outer Gas}

{\bf Distribution:} 
Eric Wilcots pointed out that dwarf irregular 
galaxies can have 
three different types of HI envelopes, which were described as 
extended smooth, extended chaotic, and starless clumps.
For example, the HI outer envelope of IC 10 has a very extended, clumpy 
counter-rotating
structure, involving a possibly infalling gas cloud of 10$^7$ M\solar.

David Strickland showed some results from detailed feedback models of 
SN-produced
outflows, mostly for larger disk galaxies, but he included a discussion of the 
shell-like structures in the outer excited gas of NGC 1569.
Sally Oey described her calculations showing how ionizing photons escape from 
irregular galaxy disks, demonstrating how the rate of escape depends on the 
porosity of the interstellar medium.

The outer gas in galaxies sometimes includes bridges between galaxy 
pairs and some 
interesting new results for the Magellanic bridge was presented by Erik Muller. 
The western part of the bridge has been mapped at high resolution and is 
found to
have a chaotic and filamentary structure. There are 163 HI shells, 
similar 
to those in the SMC, but on average smaller and less energetic.
Eon-Chang Sung et al. showed what may the largest HI bridge of all: a 
130 kpc-long 
bridge between two Blue Compact Dwarf (BCD) irregulars.

Sylvie Beaulieu described the detection of HI in the outer parts of 
two dSph galaxies 
in nearby groups, and Daniela Calzetti described the ionized 
gas in starburst galaxies.

In a report on some remarkably deep searches for CO in the 
outer parts of irregular 
galaxies, Fabian Walter showed that the outer parts of some are 
rich in molecules. For 
example, IC 10 has several clouds in distant regions as well as in the
center. Its  
H$_2$ mass is $\sim10^6$ M\solar\ and the conversion factor is like the 
Galactic value.

St\'ephanie C\^ot\'e reviewed recent results on the distribution 
of dark matter in 
dIm galaxies based on Fabry-Perot 2-d velocity fields, 
showing that they are dark 
matter dominated.

{\bf Abundances:} 
Do dwarf irregular galaxies have abundance gradients such that the outer edges 
are even 
lower in metals abundance than the central areas?  The answer to 
this question was 
touched on by a few talks and posters, but there was more 
new information regarding the 
abundances in irregulars in general, especially with relation to 
the question of their 
possible primordial nature. Starburst irregulars were featured in 
two such studies. Daniela 
Calzetti showed results of HST narrow-band imaging, which provided 
line ratios in four 
starbursters.
Alessandra Aloisi revealed the results of  FUSE spectra of IZw18, which allowed 
analysis of lines of Fe, Ar, Si, O, N, C, P, and H. She revealed the 
fact that the abundances 
of species are ten times lower in HI regions 
than in HII regions. 
The data best fit multi-burst models; the gas is not primordial and IZw18 
is not in its first starburst.

\section{The Outer Stars}

{\bf Distribution:} 
Deidre Hunter reported on the distribution of the outer stars in 
two irregular galaxies, 
detecting stellar light down to values as faint as 29 magnitudes 
per square second. ``There 
is no edge yet.''
Yutaka Komiyama showed some surprising results based on a widefield camera 
attached 
to the Suburu telescope. He found that the distribution of the 
outer stars in NGC 6822 
goes far beyond the traditional borders of the optical image and is 
similar in distribution 
and extent to HI distribution.
Ralf-J\"urgen Dettmar used very deep images of disk galaxies to 
demonstrate that the 
apparent ``edges'' are actually not cut-offs, but demarcate a steeper 
exponential distribution 
of starlight.
Studying many irregular galaxies, Igor Drozdovsky concluded 
that most are thick disks.

{\bf Star Formation Histories:}
An exhaustive review by Carme Gallart included the fact that the 
outer parts of irregular 
dwarfs are dominated by an older population, but also an intermediate 
population extends 
far from the center. She showed recent remarkable new LMC data that 
support this.
Sebastian Hidalgo demonstrated the fact that the Phoenix dwarf has a 
smooth gradient of 
age outward.
Ted Wyder, reporting on an HST study of NGC 6822, showed detailed 
star formation 
histories for five regions across the bar of the galaxy, including 
two outer areas. All 
showed star formation at all times since formation, though the outer areas 
had relatively 
lower recent activity levels.
Arna Karick reported on a survey of the irregular dwarfs in the 
Fornax cluster, with the 
intriguing suggestion that star formation activity is
preferentially located on the 
``leading edges'' of some of the galaxies, which are presumably 
falling into the cluster. 

{\bf Inside the outer edges:}
Patricia Knezek reported on a study of three extreme Cen A Group dwarfs. 
Vincent McIntyre showed some impressive HI maps of  NGC 4214's HI shells. 
Some of the outer ones have star formation at their edges.
In Caroline Simpson and collaborators' study of DDO 88, it was found that the 
galaxy has an inner HI ring, neither expanding or contracting, 
while the object is 
otherwise a normal irregular galaxy.
Janice Lee did an analysis of a sample of  H$\alpha$-selected dwarfs, 
which provided an
HI mass function for galaxies, the first of its type.
Jan Palou\v s developed expanding shell models and determined the mass spectrum of 
shell fragments.
Javier Alonso-Garc\'\i a showed {\it HST} results 
for two contrasting dwarfs in the M81 
group, one with a mixture of old and young stars and one with only old or 
intermediate age stars.
Violet Taylor measured color gradients in 100 irregular and peculiar galaxies, 
finding that most 
are redder in their outer regions, in agreement with results from CM diagrams of 
nearer Irr galaxies.
Matthew Walker showed the results of measurements of over 200 stars in 
and in the 
direction of the Fornax dwarf galaxy (not an irregular at the present, 
but 
in the past). Membership is demonstrated for
160 stars according to 
their radial velocities; the velocity profile is nearly flat. 

{\bf Formation and Evolution:}
In deriving a model for the formation of low mass irregular galaxies 
($<10^8$ M\solar), 
Massimo Ricotti calculated the effects of radiative feedback in the 
early universe; his 
simulations agree with many of the observed properties of dwarfs.
In related theoretical analyses, Gerhard Hensler described extensive 
results from 
chemodynamical models, including such items as formation, 
star formation self-
regulation, chemical peculiarities, mixing time scale,  
and infall of gas.
Stephen Murray reported on numerical simulations of dwarfs forming 
near a massive galaxy.
In related simulations carried out by Chris Fragile it was shown 
that the degree of 
enrichment by supernovae depends on the smoothness of the ISM; 
if the ISM is porous,
most of the enriched gas is lost from the galaxy core.
Arif Babul nicely demonstrated how increasingly-complete modeling is 
showing details 
of dwarf galaxy evolution, including how SN-driven outflows from dwarfs 
can pollute 
the IGM.
Sergey Mashchenko demonstrated a model for dwarf galaxies in which the 
UV radiation 
from spirals ionizes the ISM of orbiting dwarfs cycling through 
plane of the spiral, 
causing periodic star formation episodes.

{\bf Blue Compact Dwarfs:}
BCD galaxies as a class were examined in detail 
by several attendees. 
Walter Koprolin reported that an old stellar population could be detected 
underlying the brighter, young component. The structure of this 
underlying population was described by Nicola Caon, who said that 
most can be fit to an exponential law, including Mrk 35, 
though two in his sample could not. 
Finally, Eon-Chang Sung and his collaborators searched BCD galaxies 
for new super-star clusters and found several of these highly-luminous 
young products of rampant star formation.

\section{Related Issues}

{\bf Spiral Galaxies:}
Very deep HI maps of 10 spiral galaxy edges were described by 
Thijs van der Hulst, who 
concluded that UV metagalactic radiation may ionize the outer gas, 
but current simple 
models need refinement.
Annette Ferguson analyzed the outer disks of spirals, finding 
them to be surprisingly 
metal-rich and 5--10 Gyr old. She showed some faint M31 star counts 
that cover 30 square 
degrees and show structure at remarkably large distances.
Ren\'e Walterbos reported on a study of HII regions in M81's outer disk, 
including some 
in the vast M81 HI envelope. The  abundances correlate with distance 
from M81's center.
Stephen Odewahn reported on UV-optical colors of a large sample of galaxies, 
including 
analysis of the star formation history distributions.
Martin Bureau modeled NGC 2915's starless outer spiral structure.

{\bf Low Surface Brightness Galaxies:}
In a survey of Low Surface Brightness (LSB) galaxies, 
Gaspar Galaz and colleagues found 
that in general the star 
formation rate is low and there exist both color and age gradients; 
all have an old 
component and some show bulges in K-band images.
Michael Pohlen and R.-J. Dettmar investigated the structure of LSB spirals, 
including 
UGC 7321, finding that they have the same structure as normal surface brightness 
galaxies, except for the depressed overall brightness levels.

In other related fields, Lyle Hoffman and Ed Salpeter reported on the 
discovery of 10$^4$ M\solar\ 
``mini-HVCs,'' while Jay Lockman, also using the 
new GBT, 
described the discovery that the Galactic halo HI is resolved into clouds.
J. O'Brien showed an unusual galaxy that has a bar in a non-nucleated pure disk.
Andrew Odell demonstrated the effectiveness of a intemediate-band photometry of 
galaxies, mostly ellipticals in clusters, for determining ages and metallicities.
Finally, Javier Alonso-Garc\'\i a 
reported the excellent 
news that the first RR Lyrae variables have now been detected in the 
elliptical companion 
of M31, M32.

\section{Concluding Discussion Session}

The Workshop ended with a lively interchange on several
topics that had been raised during the course of the meeting.
The first question put before the attendees was ``To what extent
have the stars in the far outer regions of dwarf irregular galaxies 
formed there and to what extent have they migrated there from inner
regions of the galaxy?'' 
Discussion concluded that, on the one hand, the rotation of
the galaxy makes it difficult to move stars radially outward. 
On the other hand, such motions are easier if the gravitational 
potential has been disturbed. A change in the potential
can occur if the galaxy is perturbed by galaxy interactions, or
depending on the dark matter content, 
if an extreme starburst results in the gas being blown
out of the center and into the outer parts.

But, could stars form out there? Observational evidence suggests
that models for the formation of gas clouds due to
gravitational instabilities in the disk
fail in irregulars generally and especially so in the outer parts.
It was suggested that the thermal phase of the gas might be more important,
but such models also generally fail for dIm disks. 
It was suggested that gas disk instabilities suggested to explain 
the formation of holes seen in the outer HI disks of some irregulars 
could result
in interstices with higher column densities that might be able to form stars.
Another idea was that the outer gas could be, 
for whatever reason, in gas clouds too small 
to be detected in current interferometric observations, and stars
might be forming without massive stars so that star-forming regions
are not detectable in current H$\alpha$ surveys.
Since some extended HI disks around irregulars are found to be
highly structured and others are very smooth, it was suggested
that there might be a difference in the star formation in the
outer regions of these two sets of galaxies. Perhaps this difference
could be seen in the surface brightness profile or the colors
in the extreme outer parts of the galaxies.

The second question posed was ``What are the parallels between the
outer parts of spirals and irregulars,
and between these regions and LSB galaxies?''
The outer parts of spirals are still more metal-rich than
most irregulars.  While magnetic pressure may dominate the ISM pressure
in the outer parts of spirals, less is known about the magnetic field
in irregulars.  Spirals also show a strong decrease in the fraction of
cold neutral medium vs warm neutral medium outside the star-forming disk.
We were reminded that spirals have more shear due to differential
rotation in their outer parts than do irregulars.
Azimuthally-averaged surface brightness profiles of both LSB
and dIm galaxies show a break in the slope, with the outer
parts declining in surface brightness more steeply than the
inner parts; the cause of this break is not known. 
The outer parts of both spirals and irregulars show low-density gas
with faint populations of old and intermediate-age stars.

A third question that was raised was ``Is there any real evidence
for significant accretion onto galactic disks?'' Accretion is predicted
by cosmological and chemical evolution models, and is proposed as
the source of X-ray emission around galactic halos.
IC 10, for example, is interacting with a gas cloud that is believed 
to be falling into it.  High velocity clouds are seen around our Galaxy,
but massive ones have not been found around other galaxies.
The X-ray emission convincingly appears to result from massive star feedback.
The participants were hard pressed to think of solid observational
evidence for significant external accretion.


\medskip

\noindent
{\it Paul Hodge\footnote{University of Washington}, 
Deidre Hunter\footnote{Lowell Observatory}, 
Sally Oey$^3$}



\begin{references}
\reference{} Sandage, A. 1962, in Problems of Extragalactic Research,
ed. G. C. McVittie (New York: Macmillan Co.), p 359
\end{references}
\end{document}